\pdfoutput=1 
\documentclass{aa}
\usepackage{graphicx}
\usepackage{txfonts}
\usepackage{color}
\usepackage{tabularx} 
\usepackage{adjustbox}
\usepackage{float}
\usepackage{dblfloatfix} 
\newcommand{\be}{\begin{equation}}
\newcommand{\ee}{\end{equation}}
\newcommand{\bea}{\begin{eqnarray}}
\newcommand{\eea}{\end{eqnarray}}

\begin{document}

\title{Transient brightenings in the quiet Sun detected by ALMA at 3 mm 
}
\author{A. Nindos \inst{1}
\and C.E. Alissandrakis \inst{1}
\and S. Patsourakos \inst{1}
\and T.S. Bastian\inst{2}}
\institute{Physics Department, University of Ioannina, Ioannina GR-45110,
Greece\\
\email{anindos@uoi.gr}
\and
National Radio Astronomy Observatory, 520 Edgemont Road, Charlottesville VA
22903, USA}

\date{Received: Accepted:}

% \abstract{}{}{}{}{} 
% 5 {} token are mandatory
 
  \abstract
  % context heading (optional)
  % {} leave it empty if necessary  
{}
  % aims heading (mandatory)
{We investigate transient brightenings, that is, weak, small-scale episodes of 
energy release, in the quiet solar chromosphere; these episodes can provide insights 
into the heating mechanism of the outer layers of the solar atmosphere.}
  % methods heading (mandatory)
{Using Atacama Large Millimeter/submillimeter Array (ALMA) observations,
we performed the first systematic survey for quiet Sun transient brightenings 
at 3 mm. Our dataset included images of six $87\arcsec \times 87\arcsec$ fields
of view of the quiet Sun obtained with angular resolution of a few arcsec at a 
cadence of 2 s. The transient brightenings were detected as weak enhancements 
above the average intensity after we removed the effect of the p-mode 
oscillations. A similar analysis, over the same fields of view, was performed 
for simultaneous 304 and 1600 \AA\ data obtained with the Atmospheric Imaging 
Assembly.}
% results heading (mandatory)
{We detected 184 3 mm transient brightening events with brightness
temperatures from 70 K to more than 500 K above backgrounds of $\sim
7200-7450$ K. All events showed light curves with a gradual rise and fall, 
strongly suggesting a thermal origin. Their  mean duration and maximum area 
were 51.1 s and 12.3 Mm$^2$, respectively, with  a weak preference of appearing at
network boundaries rather than in cell interiors. Both  parameters
exhibited power-law behavior with indices of 2.35 and 2.71,
respectively.  Only a small fraction of ALMA events had either 304
or 1600 \AA\ counterparts but the properties of these events were
not significantly different from those of the general population
except that they lacked their low-end energy values.  The total
thermal energies of the ALMA transient  brightenings were between
$1.5 \times 10^{24}$ and $9.9 \times 10^{25}$ erg  and  
their frequency distribution versus energy was a power law with an index of
$1.67 \pm 0.05$.  We found that the power per unit area provided by 
the ALMA events could  account for only 1\% of the chromospheric 
radiative losses (10\% of the coronal ones).}
% conclusions heading (optional), leave it empty if necessary 
{We were able to detect, for the first time, a significant number of weak 3 mm quiet 
Sun transient brightenings. However, their energy budget falls short of meeting
the requirements for the heating of the upper layers of the solar atmosphere
and this conclusion does not change even if we use the least restrictive 
criteria possible for the detection of transient brightenings.}


   \maketitle
%
%________________________________________________________________

\section{Introduction}

Episodes of energy release are ubiquitous in the solar atmosphere and
may occur in active regions, the boundaries of the quiet Sun network
cells, and even in the cell interior and coronal holes. The larger
events, called flares, occur almost exclusively in active regions. Small-scale 
events occur everywhere and all the time,  both in active regions and the 
quiet Sun, and the detection limit of these events is determined by the sensitivity as 
well as  the spatial, temporal, and spectral resolution of the instrument. 
Therefore it is not a surprise that every time a new instrument with improved  
specifications appears, new varieties of such events, sometimes differently 
termed, are reported. 

There is substantial literature on the detection of weak transient activity
starting from early Yohkoh soft X-ray  (e.g., Shimizu 1995) and Extreme
ultraviolet Imaging Telescope (EIT) extreme ultraviolet (EUV) observations (e.g.,, Benz and
Krucker 1998, Krucker and Benz 1998; Berghmans et al. 1998)  to
observations from the Transition Region and Coronal Explorer (TRACE; e.g.,  Aschwanden et
al. 2000; Parnell \& Jupp 2000), Ramaty High Energy Solar
Spectroscopic Imager (RHESSI; e.g., Hannah et al. 2008), Hinode's X-Ray
Telescope (XRT) and EUV Imaging Spectrometer (EIS; see
Hinode Review Team et al. 2019 and references therein); Atmospheric Imaging
Assembly (AIA; e.g., Joulin  et al. 2016; Ulyanov et al. 2019), Interface Region
Imaging Spectrograph  (IRIS; e.g., Peter et al. 2014;
Vissers et al. 2015; Rouppe van  der Voort et al.  2016), Focusing Optics
X-ray Solar Imager (FOXSI), and the Nuclear Spectroscopic Telescope Array
(NuSTAR; Marsh et al. 2018). Monte Carlo simulations of the statistical
signatures of very weak events were presented by Dillon et al. (2019).

Radio observations have also been used for the detection of weak
transient activity. The great sensitivity of the radio range to small
nonthermal electron  populations that are expected as a by-product of
the reconnection process have allowed radio observations to reveal the
occurrence not only of thermal events (e.g., White et al. 1995) but
also of  nonthermal events whose emission has been attributed either to
the gyrosynchrotron mechanism (e.g., Krucker et al. 1997; Gary et al.
1997; Nindos et al. 1999) or to the plasma emission mechanism (e.g., Ramesh
et al.  2010; Saint-Hilaire et al. 2013; Suresh et al. 2017).

Weak transient activity can be detected either indirectly or
directly. Indirect detections include evidence about the occurrence of
unresolved  events by analyzing asymmetries of X-ray intensity
fluctuations (e.g., Katsukawa and Tsuneta 2001; Terzo et al. 2011), 
time lags of EUV light curves of  coronal loops (e.g., Viall and
Klimchuk 2012), and possibly subtle enhancements in the blue wing of hot
spectral lines (e.g., Hara et al. 2008; De Pontieu et al. 2009; Brooks and 
Warren 2012). Direct evidence includes the detection of time
variability of the intensity (e.g., Berghmans et al.  1998) or the emission
measure (e.g., Krucker and Benz 1998)  of clusters of pixels above some
predefined threshold. In most studies, most of the detected events are
spatially unresolved (e.g., Benz and Krucker 1999), but some events may
exhibit a loop-like morphology (e.g., White et al. 1995; Benz \& Krucker 1998;
Warren et al. 2007), a jet-like morphology (e.g., Tian et al. 2014; Alissandrakis 
et al. 2015), show  evidence of two-loop interactions (e.g., Shimizu 
et al.  1994; Alissandrakis et al. 2017), or even resemble Ellerman bombs 
(e.g., Shetye et al. 2018). 
   
Several authors have studied the contribution of weak transient events
to the heating of the upper atmosphere. These studies are motivated by
the  nanoflare heating conjecture first proposed by Parker
(1988). Nanoflares, events with energies less than 10$^{24}$ erg, are
expected to occur as a  result of magnetic reconnection in elemental,
tangled magnetic flux tubes that  are below the resolution limit of
present-day instruments. So far only detections of individual events
with energy down to the ``high-end'' limit of Parker's estimate have
been achieved (e.g., Berghmans et al. 1998; Parnell and Jupp 2000;
Aschwanden et al. 2000; Winebarger et al. 2013;  R\'{e}gnier et
al. 2014; Joulin et al. 2016; Subramanian et al. 2018).  It has been
argued that these small events could heat the upper atmosphere if the
energy released during various types of flare-like activity follows a
power-law frequency distribution with energy that has an index $\alpha
\geq 2$ (e.g., Hudson 1991). Several authors have presented such
computations (e.g., Crosby et al. 1993; Shimizu 1995, Krucker and Benz
1998; Aschwanden et al. 2000; Benz and Krucker 2002).  Most studies
(with some notable exceptions, e.g., Krucker and Benz 1998; Parnell and
Jupp 2000; Benz and Krucker 2002) show that the above requirement  is
not satisfied.

Any attempt to solve the coronal heating problem cannot circumvent the
problem of determining the mechanisms by which chromospheric plasma is
heated and lifted to form the corona. The detection of ubiquitous weak
enhancements in the blue wing of hot spectral lines has been
interpreted by some authors (e.g., De Pontieu et al. 2009) as evidence
of upflows associated  with chromospheric spicules that provide a
significant mass supply mechanism  for the corona. This scenario,
which ultimately places the source of coronal heating in the
chromosphere, has been contested by Klimchuk (2012) who argued that
the preheated plasma provided by spicules is not sufficient to fill the
corona, but even if it is, this would require additional heating to remain
hot as it rises into the corona. 

No matter what the role of chromospheric spicules in coronal heating
is, the energy requirements to heat the quiet chromosphere is about
one order of magnitude higher than those for the quiet corona
(Withbroe \& Noyes 1977). Then an obvious question is to  what extent
chromospheric small-scale energy release  events contribute to the
heating of the chromosphere in general, and the  quiet chromosphere in
particular. The calculation of the physical parameters of such events
in optical and UV observations is complicated because this calculation relies on
complex  physical effects such as partial and time-dependent
ionization and departures  from local thermodynamic equilibrium
(e.g., Leenaarts et al. 2013). However, that  task can be facilitated
by using millimeter wavelength observations of the quiet  chromosphere
because the relevant emission mechanism is thermal free-free,  which
does not suffer from the problems mentioned above and because, thanks
to the Rayleigh-Jeans law, the recorded brightness temperature is
linearly linked to the electron temperature (e.g., Shibasaki et
al. 2011; Wedemeyer  et al. 2016; but see Martinez-Sykora et al. 2020
who suggest that the interpretation of millimeter-$\lambda$ continua might be
more complicated than these simple expectations).  

The availability of millimeter-$\lambda$ solar observations with Atacama Large Millimeter/submillimeter Array (ALMA) has the
potential to provide new insights into the physics of the chromosphere
thanks  to the superior spatial and temporal resolution and  sensitivity of the instrument. Observations of the quiet Sun with ALMA have
been reported by Shimojo et al. (2017a,b), White et al. (2017),
Alissandrakis et al. (2017), Bastian et al. (2017), Yokoyama et
al. (2018), Nindos et al. (2018), Loukitcheva et al. (2019), Braj\v{s}a et
al. (2018), Jafarzadeh et al.(2019), Selhorst, et al. (2019), Molnar et al.
(2019), Patsourakos et al. (2020), and Wedemeyer et al. (2020). Among those 
articles, detection of single weak transient events was reported by 
Shimojo et al. (2017b), who studied a plasmoid ejection associated with an 
X-ray bright point, and by Yokoyama et al. (2018) who reported jet-like 
activity at 3 mm. 

In this article we present the first systematic survey for weak
small-scale  energy release events in the quiet Sun using ALMA data
obtained at 3 mm. Different authors use different terminology for such events;
in what follows we adopt the term ``transient brightenings''. Our 
article is structured as follows: In Section 2 we discuss the observations
and our analysis. The statistics and properties of the detected transient
brightenings are given in Section 3 while their implications for chromospheric
heating are discussed in Section 4. Finally, we present conclusions in Section 
5.

\section{Observations and data analysis}

We used the ALMA observations of the quiet Sun presented by Nindos et
al. (2018), which included seven 120$\arcsec$ circular fields of view
(targets), observed at 100 GHz (3 mm) on March 16, 2017. These
targets, numbered from 1 to 7, correspond to $\mu =
[0.16,0.34,0.52,0.72,0.82,0.92,1.00]$  along a position angle of
135\degr\ from  the center-North direction and thus supplied a
center-to-limb coverage ($\mu = \cos \theta$, where $\theta$ is the
angle  between the line of sight and the local vertical). For our
analysis we  used only targets 2-7 because a significant part of the
target 1 field of view contained off-limb locations. Each target was
observed for 10 min with a cadence of 2 s. The reduction of the
ALMA data has been described by Nindos et al.  (2018). The final
images had pixel size of 1$\arcsec$ and their spatial resolution
deduced from the clean beam size was about $2.5\arcsec \times
4.5\arcsec$ with the exception of target 7 whose resolution was
$2.3\arcsec  \times 8.1\arcsec$. 

In our analysis we used the central $87\arcsec \times 87\arcsec$
region of  each field of view to avoid artifacts introduced by the
primary beam correction toward the edge of the images. Since the
characteristic scale of the  chromospheric network is about
20$\arcsec$-30$\arcsec$, the regions we selected contain an adequate 
number of supergranules to allow us to perform meaningful statistics of 
the occurrence of 3 mm transient brightenings in the quiet Sun.

We also analyzed AIA images (same time intervals and fields of view as
the ALMA images) obtained at 304 \AA\ and 1600 \AA, primarily to compare these  with the 3 mm results. The cadence of the AIA images
was 12 s in 304 \AA\ and 24 s in 1600 \AA. The analysis of the AIA
data involved the correction for differential rotation, their
convolution with the appropriate ALMA beam, and their co-alignment with
the ALMA images by cross-correlating the time average images for each
target. We note that the AIA 304 \AA\ channel records primarily emission from the
upper chromosphere, while the emission in 1600 \AA\ partly comes from
the upper photosphere.

\begin{figure}[!h]
\centering
\includegraphics[width=0.50\textwidth]{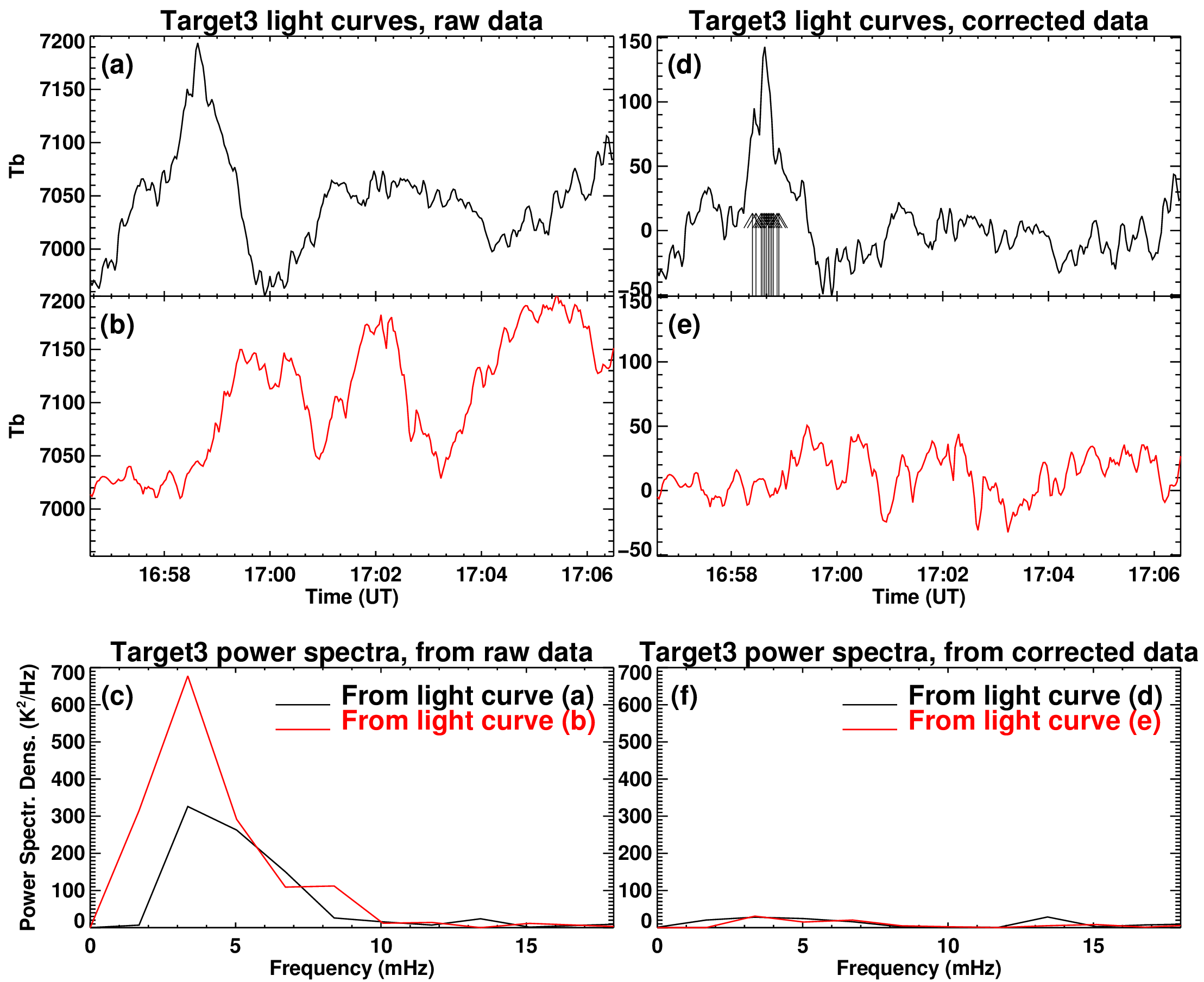}
\caption{Top row: Light curves of a target 3 single pixel showing transient 
brightening before (a) and after (d) the processing described in 
section 2. In this and subsequent light-curve displays, the arrows indicate 
the times in which the intensity exceeds the 2.5$\sigma$ threshold above average. 
Middle row: Same as top row for a pixel that did not show transient 
brightening. The curve in panel (b) has been displaced by -300 K. Bottom row: 
Panel (c) shows the power spectra deduced from light curves (a) and (b), while 
panel (f) shows the power spectra deduced from light curves (d) and (e).}
\end{figure}

For the detection of transient brightenings we used the following
method.  For each target we computed the light curve of each
pixel. Each light curve was detrended by subtracting the running
average of the time profile of a macropixel with size equal to the
ALMA beam, which was centered around the pixel under consideration.
The running average was computed with a kernel of 10 min (i.e., the
width of the  kernel was equal to the duration of observations of each
target). Detrending the light  curves was necessary in order to remove
long-term, slowly varying trends from the data,  which corresponded to
neither oscillations nor transient brightening activity. The use  of
macropixels for the background subtraction was deemed necessary for
the suppression of artifacts that may appear when a single
pixel is used for the background calculation (see Berghmans et al. 1998).

Then an attempt was made to remove the effect of  oscillations from
the data since it is well known that p-mode oscillations are
ubiquitous in $\sim$3 mm quiet Sun data (see White et al. 2006). Recently,
Patsourakos et al. (2020) analyzed such oscillations using the same
dataset as ours and found that the frequency of oscillations was $4.2
\pm 1.7$ mHz with an rms of about 55-75 K (i.e., approximately up to 1\% of the
recorded averaged brightness  temperatures), while their amplitude in
individual pixels can reach values as high as 350 K. 

In our detrended light curves we applied a least-squares curve fitting
procedure (see Roerink et al. 2000) based on the harmonic components
that   correspond to the oscillations detected by Patsourakos et
al. (2020).  The  amplitude and phase of these functions were
determined iteratively by  removing data points with large deviations
from the fitting curve. The  remaining points were used for the
recalculation of the coefficients until an acceptable maximum error
($\lesssim 5$ K) is reached or their number becomes less than
five. The final sinusoidal curves were subtracted from the  detrended
light curves. Two examples of light curves of individual pixels
appear in Fig. 1; panels (a) and (b) show the original light curves
while panels (d) and (e) shows the light curves after the application of
the processing described above. 

The bottom row of Fig. 1 shows the power spectra deduced from the
light curves  presented in the same figure. Prior to the calculation
of the power spectra of panel (c) the light curves were detrended
while the light curves of panels (d) and (e) were submitted to power
spectral analysis without any additional processing. Panel (c) shows
that the p-mode oscillations stand out prominently in the power
spectra resulted from the uncorrected light curves while panel (f)
indicates that our algorithm can significantly suppress the power of
p-mode oscillations (by factors of 12 to 17). We note that such
behavior was typical for all pixels. The maximum residual p-mode power
is about 30  K$^2$/Hz in the examples of Fig. 1 and can reach values
as high as 40 K$^2$/Hz with a slight increase toward the limb. That
trend could affect the detection of very weak events, and we return 
to that point in Section  3.

In the corrected light curves, an ``event'' was identified if: (i) the
intensity of at  least four consecutive points of the light curve was
above some user-defined  threshold above the average intensity of the
curve, (ii) such behavior  was also exhibited by a user-defined number
of clusters of pixels adjacent to the pixel under consideration, and
(iii) a synchrony tolerance of $\pm 2$ min between light curve peaks
of the selected adjacent pixels was satisfied (see Benz \& Krucker 2002
and references therein).  
%\LEt{Quotation marks can be used to introduce a special meaning for a word or phrase the first time it is used in the text, but thereafter quotation marks are not needed. If the special
%meaning is otherwise clear, or indicated by "so called" or similar, the quotation
%marks are not needed. Please amend throughout.}
These criteria were implemented to avoid the
accidental  inclusion of exceptionally large statistical fluctuations
or sidelobes. 

Our choices certainly affect the statistics of the detected events and
this point is discussed in section 3. Most of the subsequent results
have been derived using a multiplication factor of 2.5$\sigma$ above
average intensity for the light curves (that is a threshold that has
been employed in several previous studies, e.g.,  Berghmans et
al. 1998) and a beam-size cluster of adjacent pixels for the spatial
criterion (i.e., 12 pixels for targets 2-6 and 19 pixels for  target
7). The latter selection is justified by the fact that due to the
limited spatial resolution, the intensity of individual pixels should
be correlated on the scale of the synthesized beam. In Fig. 1 the
top row  of light curves belong to a pixel that  was part of one of the
events selected  by our method and the arrows point to  intensities
which exceed the 2.5$\sigma$ threshold. On the other hand, the
middle row of light curves belong to a pixel that did not show transient
activity.

We followed the same procedure for the identification of  transient 
brightenings in the AIA (see Lemen et al. 2012) 304 \AA\ and 1600 \AA\ 
data with the exception that we required one light curve point above the 
2.5$\sigma$ threshold instead of four consecutive points due to the lower 
cadence of the AIA data. 

\section{Statistics and properties of transient brightenings}

%The number of ALMA events per target are given in the second column of Table 1.

\begin{figure}[t]
\centering
\includegraphics[width=0.50\textwidth]{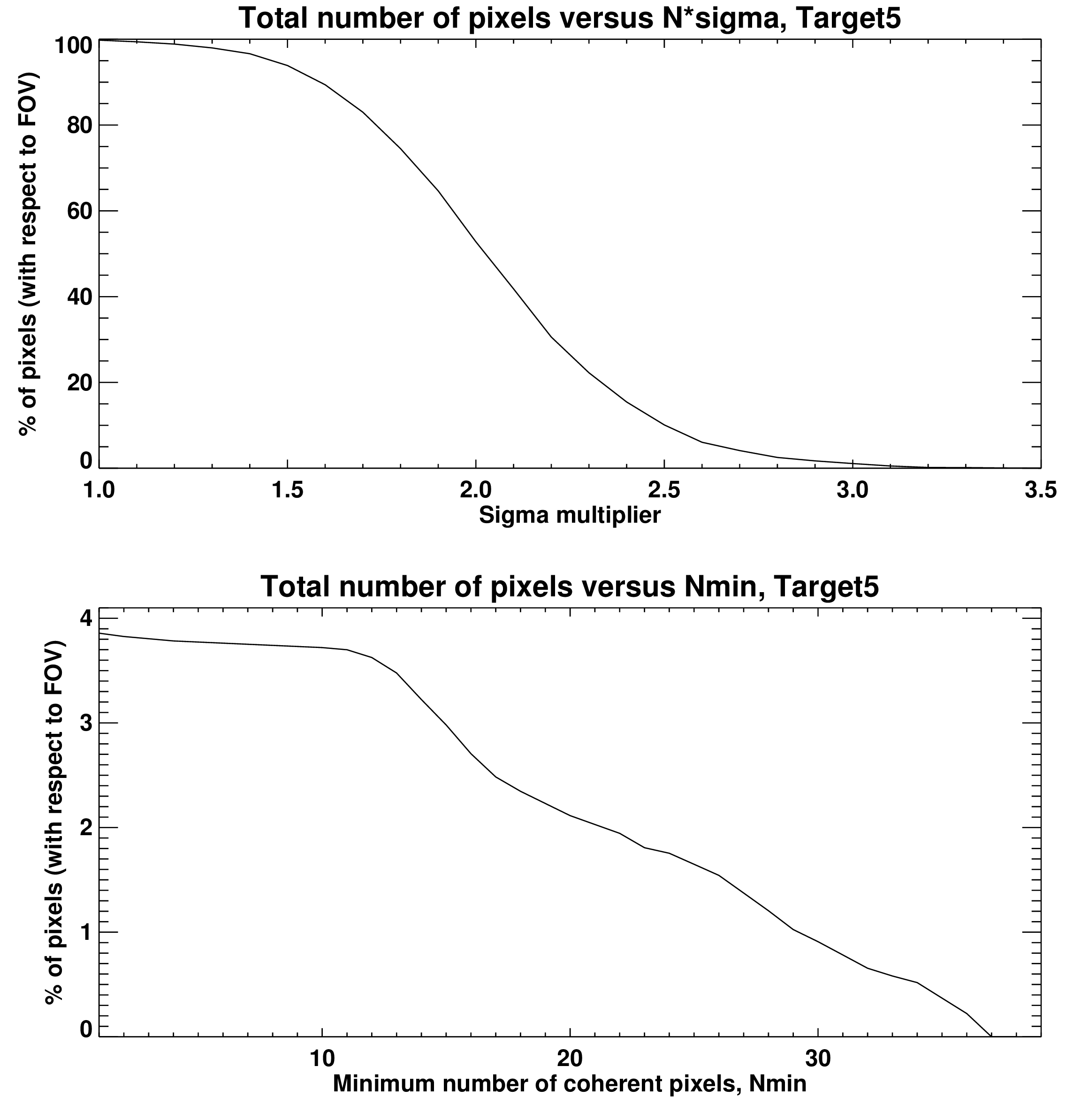}
\caption{Top: Ratio of the target 5 ``event pixels'' with respect to the number
of pixels of the whole field of view (expressed in percentage) as a function of 
the $\sigma$ multiplication factor that determines the threshold for the 
detection of transient brightenings. No additional spatial coherence 
detection criteria have been applied for the production of these results.
Bottom: Same quantity as top curve but now computed as a function of the 
minimum number of adjacent pixels on top of the 2.5$\sigma$ 
multiplication factor.}
\end{figure}

\begin{table*}[h]
%{\small
\begin{center}
\caption{Statistics of transient brightenings}
\label{ALMAobs}
\begin{tabular}{lcccccc}
\hline 
Region & $\mu$ & ALMA TBs & 304 \AA\ TBs & 1600 \AA\ TBs & ALMA TBs with & ALMA TBs with  \\
       &     &     &              &               & 304 \AA\ counterparts & 1600 \AA\ counterparts \\
\hline
Target 2 & 0.16 & 23 (2.4\%)  & 36 (5.7\%) & 119 (15.2\%) & 3 & 3 \\
Target 3 & 0.34 & 30 (3.8\%) & 35 (5.0\%)  & 117 (18.1\%) & 2 & 3 \\
Target 4 & 0.52 & 32 (5.4\%) & 33 (5.6\%) &  109 (17.5\%) & 4 & 2 \\
Target 5 & 0.72 & 53 (6.9\%) & 31 (6.9\%) & 111 (15.4\%) & 4 & 2 \\
Target 6 & 0.82 & 34  (4.9\%) & 35 (6.4\%) & 91 (15.1\%) & 4 & 2 \\
Target 7 & 1.00 & 12  (1.7\%) & 29 (6.1\%) & 86 (13.4\%) & 1 & 2 \\
\hline 
\end{tabular}
\tablefoot{Values in parentheses indicate percentages of the event pixels
with respect to the pixels of the whole field of view.}
\end{center}
%}
\end{table*}

\begin{table*}
%{\small
\begin{center}
\caption{Duration, size, and location of transient brightenings}
\begin{tabular}{lccc}
\hline 
Parameter                 & 3 mm  & 304 \AA\ & 1600 \AA\  \\
\hline
Network events (all events) & 68\% & 69\% & 43\% \\
Network events (paired events) & 72\%\tablefootmark{a} & 75\%\tablefootmark{b} & 51\%\tablefootmark{c} \\
Mean area (Mm$^2$) (all) & $12.3 \pm 3.4$ & $14.3 \pm 6.0$ & $15.5 \pm 6.7$ \\
Power-law index of distribution (all) & $2.73 \pm 0.02$ & $2.37 \pm 0.03$ & $2.32 \pm 0.02$ \\
Mean area (Mm$^2$) (paired)  & $(13.1 \pm 3.9)$\tablefootmark{a} & - & - \\
Power-law index of distribution (paired) & ($2.71 \pm 0.03$)\tablefootmark{a} & - & - \\
Mean duration\tablefootmark{d} (s) (all) & $51.1 \pm 6.5$ & $52.7 \pm 5.8$ & $63.4 \pm 5.6$ \\
Power-law index of distribution (all) & $2.35 \pm 0.02$ & $2.65 \pm 0.01$ & $3.11 \pm 0.01$ \\
Mean duration\tablefootmark{d} (s) (paired)  & $(53.1 \pm 13.2)$\tablefootmark{a} & -   & - \\ 
Power-law index of distribution (paired) & ($2.31 \pm 0.02$)\tablefootmark{a} & - & - \\
%\\
\hline 
\end{tabular}
\tablefoot{No values related to the area and duration are 
reported for the AIA events at a given passband that had ALMA counterparts owing
to their small number. 
\tablefoottext{a}{3 mm events paired with either 304 \AA\ or 1600 \AA\ events.}
\tablefoottext{b}{304 \AA\ events paired with 3 mm events.}
\tablefoottext{c}{1600 \AA\ events paired with 3 mm events.}
\tablefoottext{d}{FWHM.}
}
\end{center}
%}
\end{table*}

In Fig. 2 we show how the change of our detection thresholds affects
the statistics of transient brightenings in target 5. Similar results
were obtained  for the other targets. In the top panel we show the
influence of the  change of the detection threshold above background
in the light curves of individual pixels that resulted from the
application of our method without imposing any additional  spatial
coherence detection criterion. Practically all pixels show intensity
variations within the 1.0-1.2$\sigma$ level and this reflects the
presence of noise and/or the influence of oscillations that cannot be
removed completely from  the light curves (see also the discussion
about Fig. 1). The percentage of selected pixels drops quickly for
thresholds higher than  1.7$\sigma$, exhibits an inflection point
around 2.6$\sigma$, and approaches zero for thresholds  higher than
3.0$\sigma$.

In the bottom panel of Fig. 2 we show how the transient brightenings
statistics are affected when we vary the event ``spatial coherence''
criterion (i.e., the minimum number of adjacent pixels, $N_{min}$) 
on top of the 2.5$\sigma$ multiplication factor. In 
the two plots of Fig. 2 the percentages are different; in the top panel 
the value that corresponds to
2.5$\sigma$ is $\sim 10\%$, while in the bottom panel all values are
less than 4\%. This difference in the percentages is because an
additional spatial coherence criterion ($N_{min}$ coherent pixels)
was imposed on top of the 2.5$\sigma$ threshold for the bottom
curve.  The number of
selected event pixels shows a plateau for $N_{min} \leq 12$, which is
the number of pixels that correspond to  the beam size for targets
2-6. The plateau appears to result from the limited spatial resolution of the
images, which yields the correlation of signals from individual point
sources on map patches containing 12 pixels (or 19 pixels for target
7). Furthermore, the lack of appreciable changes for $N_{min} \leq 12$
justifies our selection of using $N_{min} = 12$ in the subsequent
analysis. The percentage of selected event pixels drops rather
smoothly for $N_{min} > 12$; the half maximum value is reached for
$N_{min} = 23$ while there is no event with more than 37 pixels. 

\begin{figure}[!h]
\centering
\includegraphics[width=0.50\textwidth]{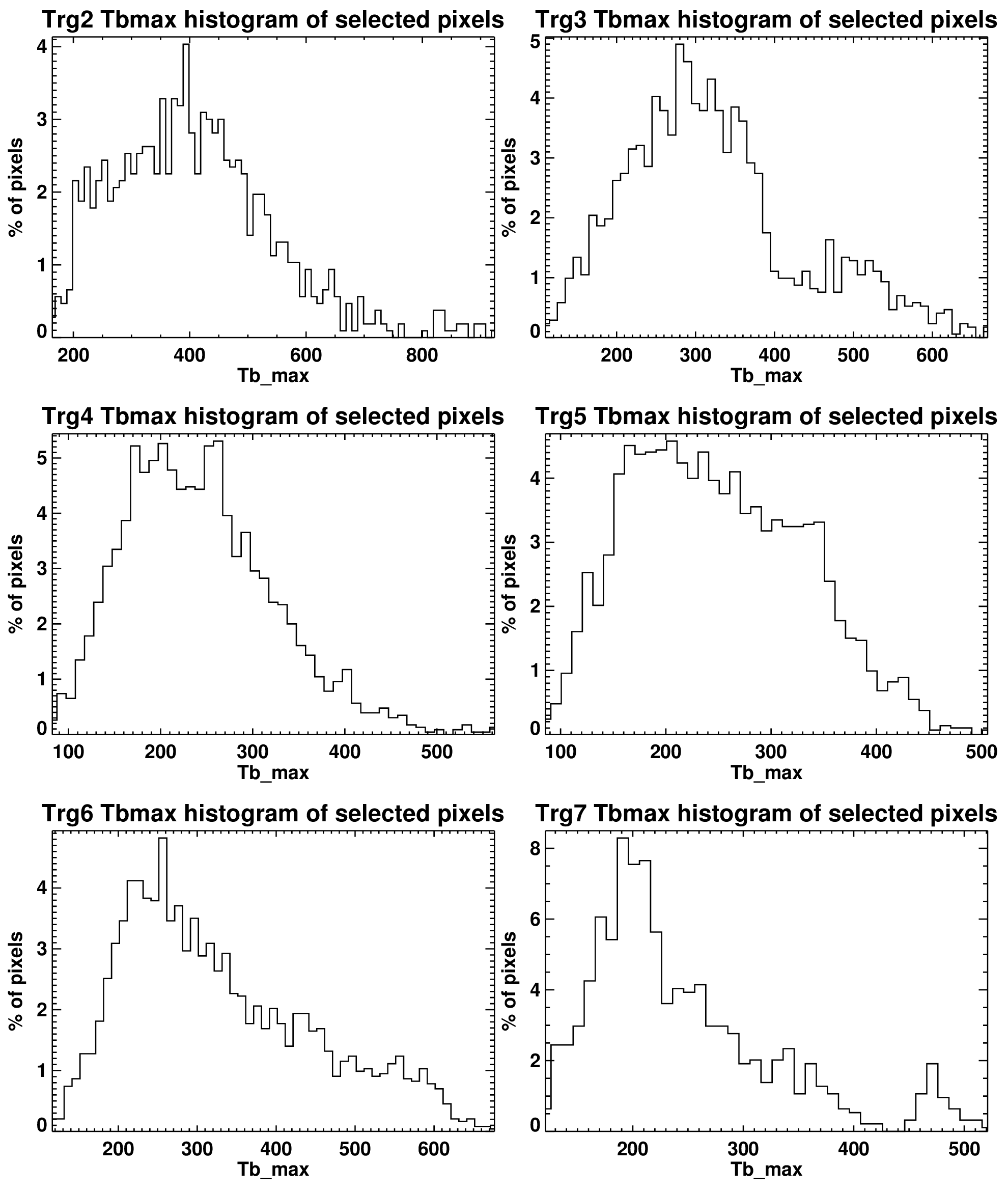}
\caption{Histograms of maximum brightness temperature above the background 
of the selected event pixels for all targets.}
\end{figure}

\begin{figure}[!h]
\centering
\includegraphics[width=0.50\textwidth]{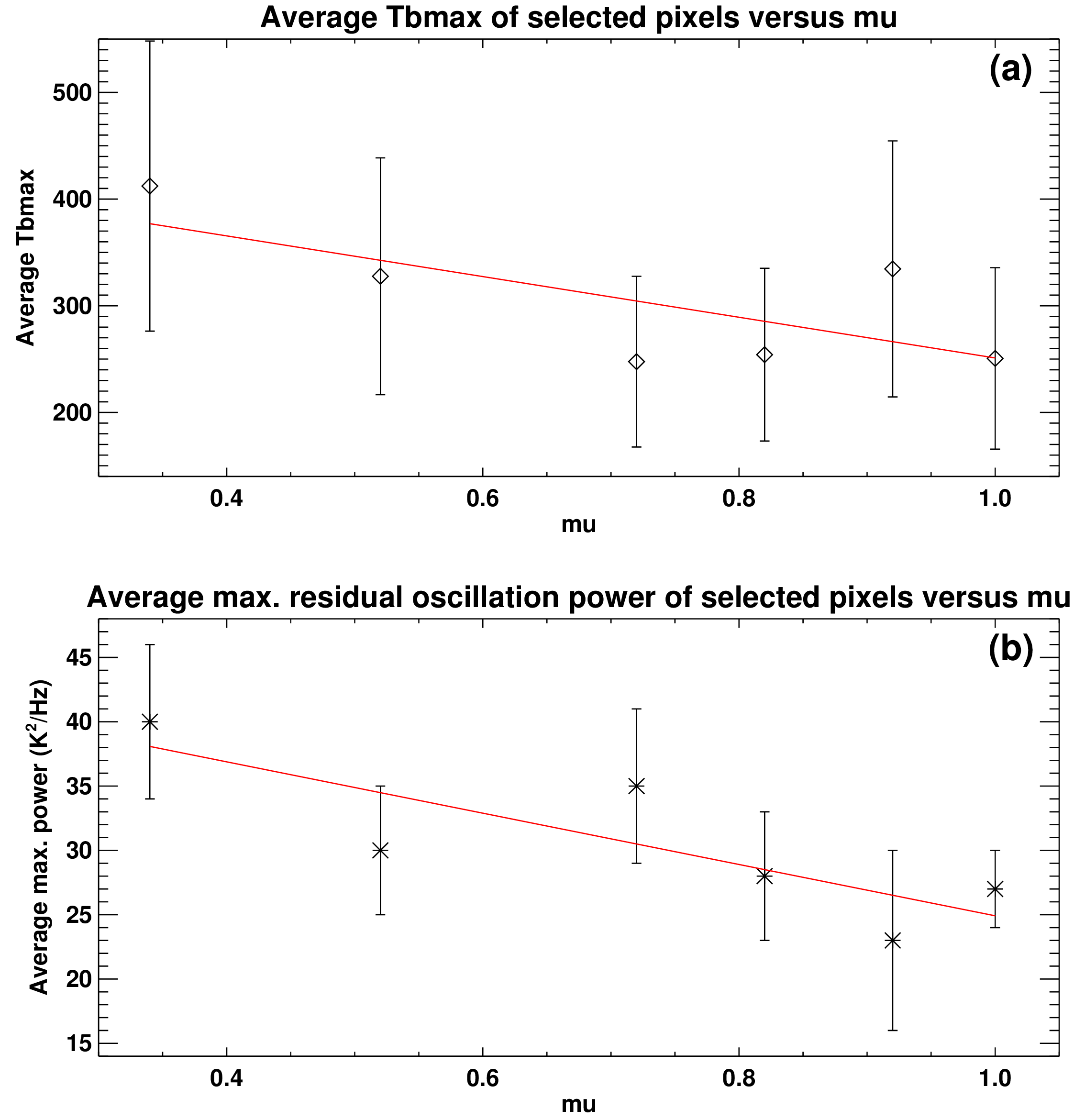}
\caption{Top: Average maximum brightness temperature of the selected event pixels 
in each target as a function of $\mu$. The error bars denote the standard 
deviation of the corresponding distributions. The red line shows the linear-squares fit to the measurements. Bottom: Same as top, but for the average maximum
residual oscillation power.} 
\end{figure}

The number of ALMA events per target that were detected after we
applied our  method with the 2.5$\sigma$ criterion for  the light
curves and the beam-size patch of adjacent pixels for the spatial
criterion are given in the third  column of Table 1. For each target,
we also give the percentages of event  pixels with respect to the
pixels of the whole field of view. 

In Fig. 3 we present (for all targets) histograms of the maximum
brightness  temperature above the background of the pixels that
exhibited transient  brightenings.  All histograms show a rise part 
that is steeper than their decay part, which probably indicates that
the populations of events with relatively high intensities are larger
than the populations of smaller events.  Equivalently, this conclusion
indicates departure from Gaussianity for the maximum brightness
temperature distributions, which is further supported by the analysis
that is presented in this section as well as in Section 4. We also
note that the histograms of targets 3, 5, and 6 appear bimodal, which could
suggest the existence of two populations of event pixels regarding their
maximum brightness temperature.  The histogram widths for targets 2-6, 
measured at half maximum, are between 110 and 230 K.  The histogram width 
for target 7 is about 60 K, much more narrow than the others, and that is 
probably a spatial resolution effect (see Section 2 and Nindos et al. 2018).

The histograms of Fig. 3 provide hints for center-to-limb variation of
the  average maximum brightness  temperature, $T_{b,max}$, which
appears to slightly increase toward the limb.  This effect is better represented in Fig. 4(a), in which we plot that quantity as a function  
of $\mu$. The scatter that appears in this figure is significant, however
it is  possible to obtain a linear least-squares fit of the data within
the error bars. The center-to-limb variation of $T_{b,max}$ is 
consistent with the slight increase of residual oscillation power toward the 
limb  indicated by Fig. 4(b): for an event detection, 
$T_{b,max}$ needs to increase  to compensate for the increase of the 
residual oscillatory power. Again the scatter of data points in Fig. 4(b) is 
significant but the limbward increase of residual power can easily be 
identified as evidenced by the least-squares fit to the data points. It is 
possible that the tendency reflected in Fig. 4(b) is due to the moderate 
increase of the 3 mm p-mode oscillations toward the limb (see Patsourakos et 
al. 2020).

There is an increase in the number of detected events toward disk 
center in targets 2-5 (see third column of Table 1). This trend is not shared 
by the number of events detected in targets 6 and (especially) 7 possibly owing
to inferior seeing conditions during the observations and larger 
beam size (target 7). With other conditions (e.g., seeing and spatial 
resolution) remaining the same, we would expect fewer detections as $\mu$
decreases owing to the increased degree of obscuration due to  spicules,
and this is the case for targets 2-5.

A complementary approach to studying the distribution of
the transient  brightenings occurrence rate as a function of height in more detail could
be provided by the analysis of snapshot maps made in different
spectral windows (spws): for example,  spw 0 (93 GHz) versus spw 3 (107
GHz). Such approach was followed by Rodger et al. (2019) for the
analysis of a plasmoid ejection. However, in our  analysis we summed
over all spectral windows when producing the snapshot maps  and
therefore the results from such computations could be addressed in
a future work.    

\begin{figure*}[t]
\centering
\includegraphics[width=0.8\textwidth]{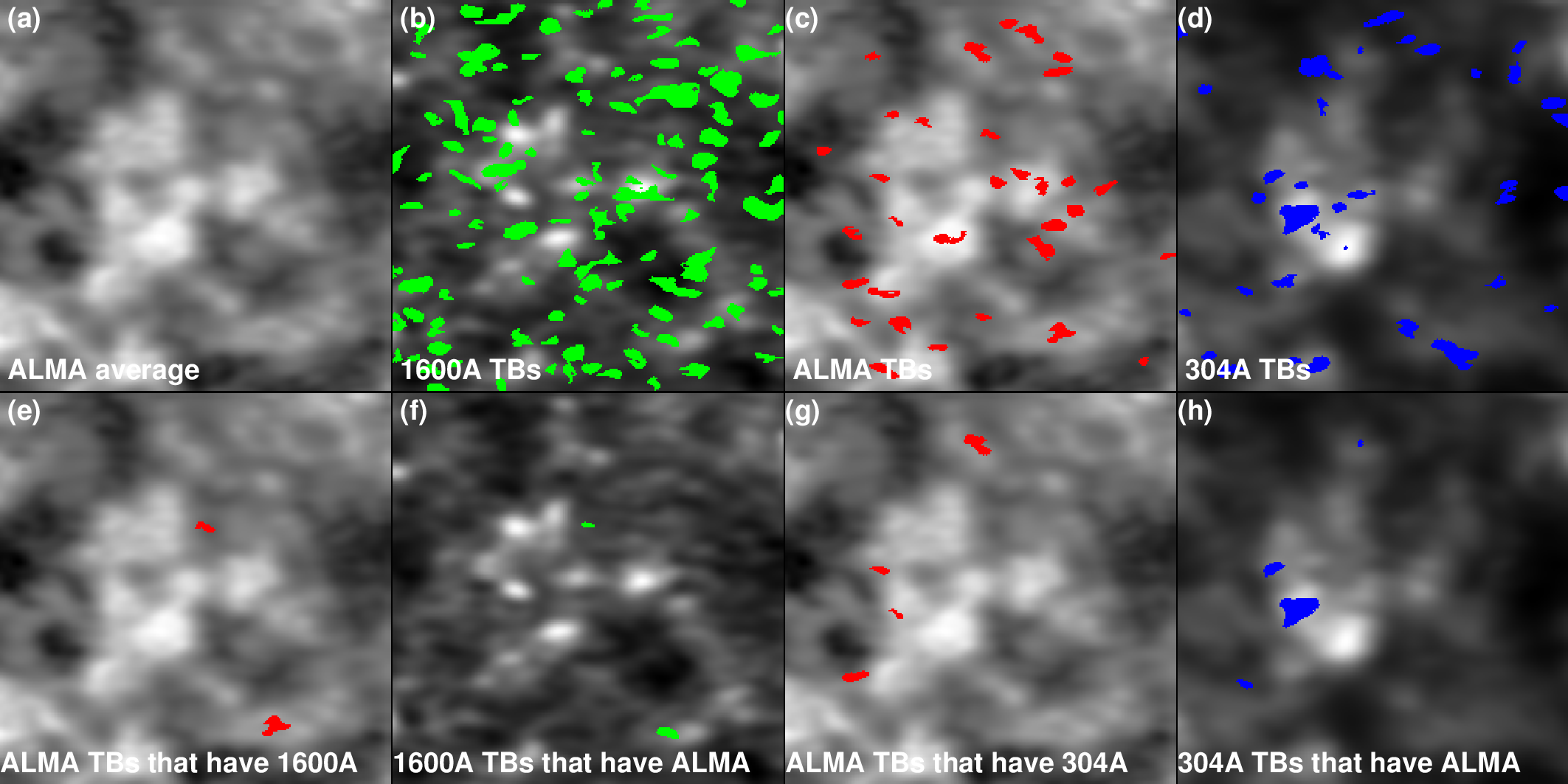}
\caption{Display of the selected pixels that correspond to transient 
brightenings in 1600 \AA\ (green), 3 mm (red), and 304 \AA\ (blue) target 4 
data. In the top row all events detected at a given wavelength are denoted. In 
the bottom row, panels (e) and (f) show only the events that appear both at 3 mm
and 1600 \AA, respectively, while panels (g) and (h) show only the events that 
appear both at 3 mm and 304 \AA, respectively. In each panel the relevant
average image is shown as background.}  
\end{figure*}

\begin{figure*}
\centering
\includegraphics[width=0.8\textwidth]{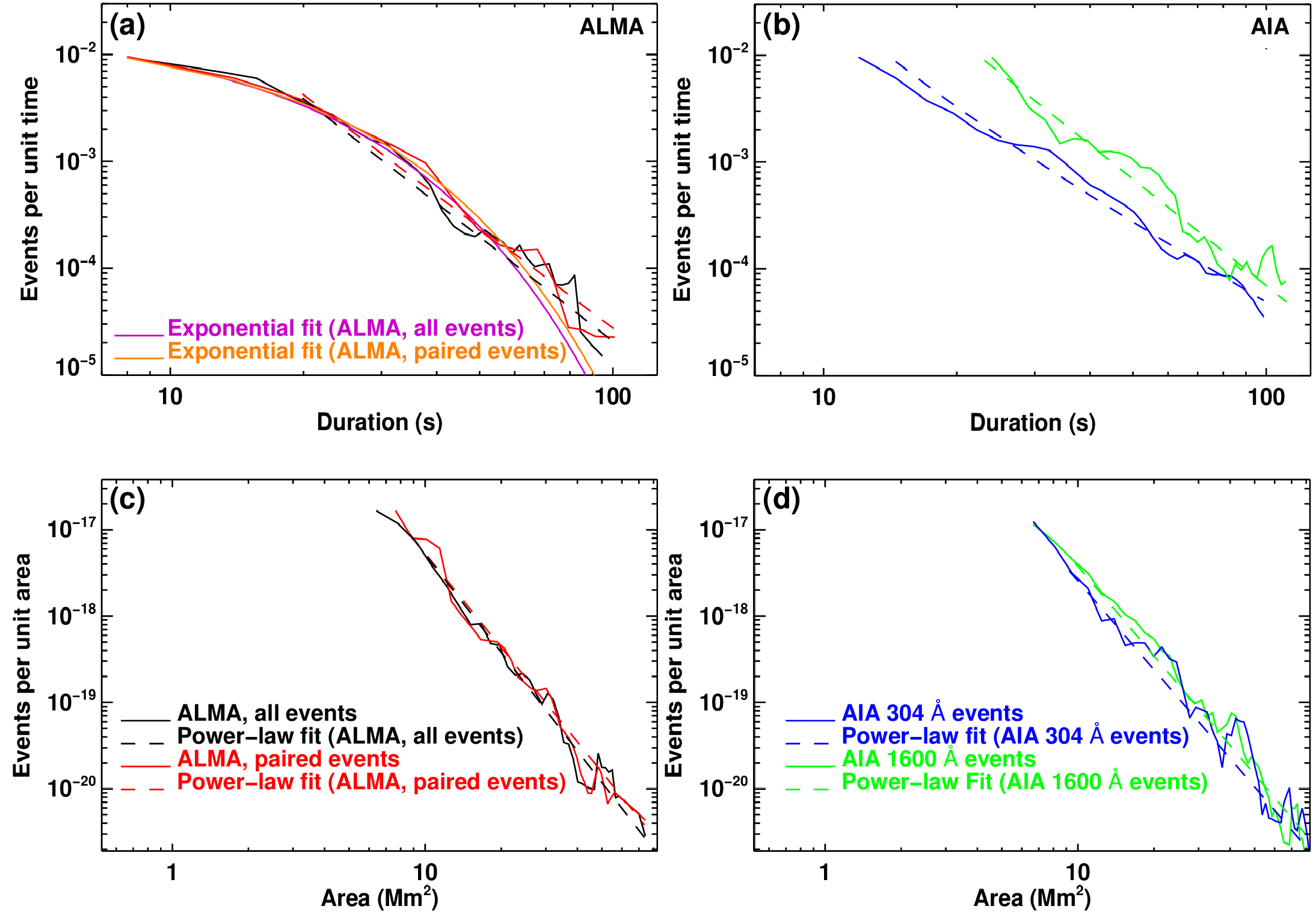}
\caption{(a) Frequency distribution of duration for all events detected
by ALMA (black solid curve) and for those with AIA counterparts, either 304
or 1600 \AA, (red solid curve). The dashed black and dashed red lines 
represent the fitting of those distributions with power-law functions,
while the purple and yellow curves indicate their fitting with
exponential functions. (b) Frequency distributions of duration for all events 
detected at 304 \AA\ (solid blue) and 1600 \AA\ (solid green) and their 
fittings with power-law functions (dashed blue and dashed green, respectively).
(c) and (d) Same as (a) and (b), respectively, but for the frequency 
distribution of maximum area.}  
\end{figure*}

For both AIA 304 \AA\ and 1600 \AA\ datasets we detected transient
brightenings as described in Section 2. The number of AIA events per
target,  together with the percentages of event pixels with respect to
each field of  view, are given in the fourth (304 \AA) and fifth (1600
\AA) columns of Table  1. The total number of ALMA events (184) is almost
equal to the total numbers of 304 \AA\  events (199), but it is a factor
of 3.4 smaller than that of the 1600 \AA\ events. 

An interesting question is how the ALMA transient brightenings correlate with 
the AIA brightenings. In order to find possible counterparts of the ALMA events among 
those detected in the AIA datasets (either 304 or 1600 \AA) we requested that the two events (ALMA and AIA) have at least one common pixel and that their
light curves show maxima within 15 s. We note that the cadences
of the 3 mm, 304 \AA,\, and 1600 \AA\ data were 2, 12, and 24 s, respectively.
%\LEt{Please check for intended meaning.} 
The
synchrony tolerance that we adopted was about 23-29\% of the full width at half maximum (FWHM) of 
the average duration of the ALMA and AIA events (see Table 2). The number of 
paired ALMA-304 \AA\ events and ALMA-1600 \AA\ events appear in columns 6 and 
7 of Table 1. Their number is much smaller than the number of the events 
detected independently in each database; the percentage of their pixels with
respect to each field of view have not been included in Table 1 because they 
are all below 0.5\%. We also note that we found no event 
appearing in all three datasets. If we increase the accepted temporal 
difference between maxima of the ALMA and AIA events to the average of the 
FWHM values of Table 2, the number of paired events roughly doubles. 

We also estimated how many ALMA transient brightenings with AIA 
counterparts would be expected if there was no correlation between ALMA and 
AIA events, and any concurrence would be by pure chance. To this end, for each
target we used the percentages of the field of view occupied by the events (see
the values in parentheses in columns 3-5 of Table 1) to find the number of 
pixels that could be part of paired events by pure chance. The result was 
divided by the mean number of pixels of the ALMA events of a given target.
The number of chance concurrences ranged from less than 1 to 3 events and was
always smaller than the number of detected paired events with the exceptions
of target 4 and 5  ALMA-1600 \AA\ paired events.

\begin{figure*}[t]
\centering
\includegraphics[width=1.00\textwidth]{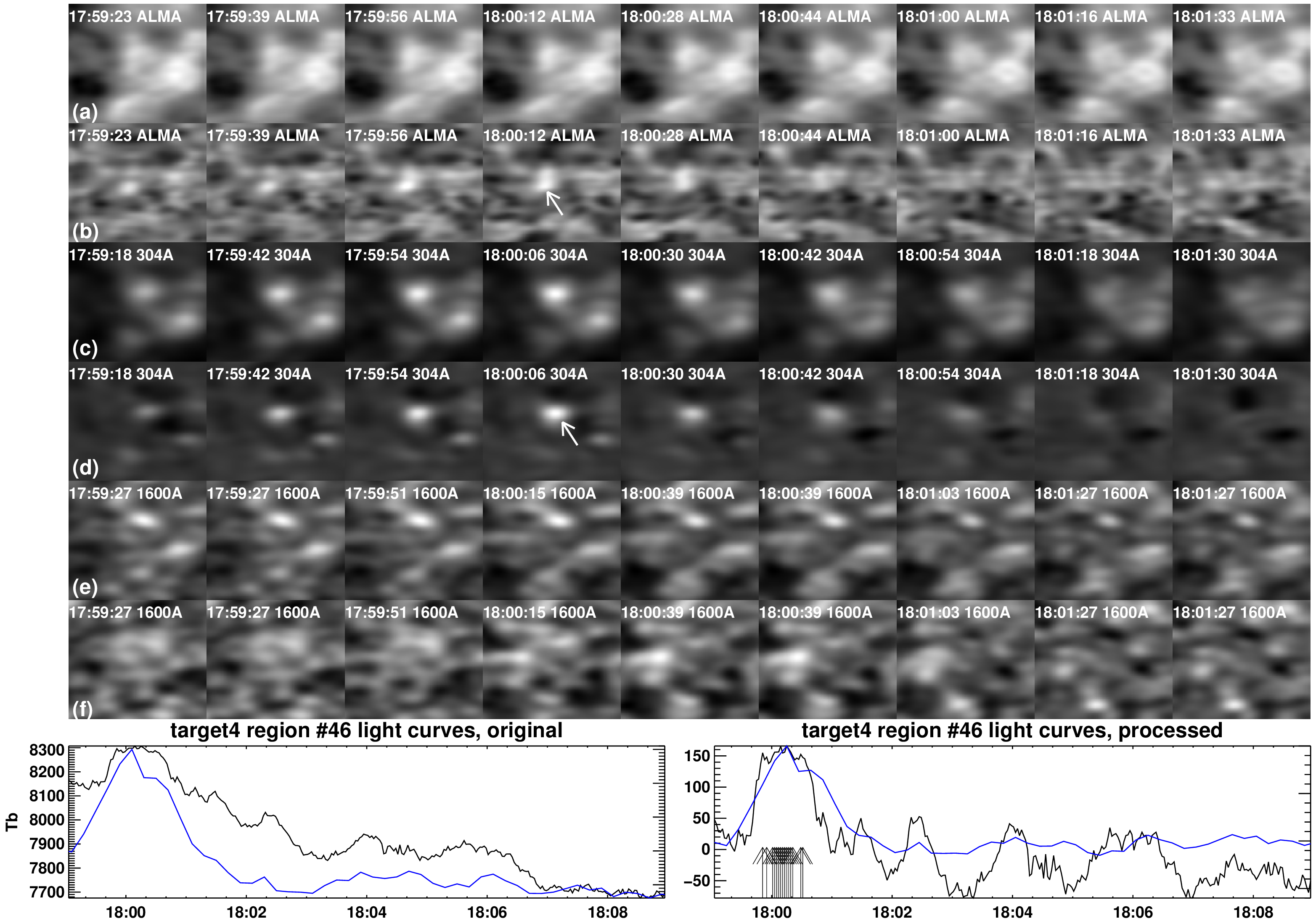}
\caption{A transient brightening detected both in 3 mm  and 304 \AA\ 
target 4 data. Row (a) shows characteristic ALMA snapshots with field of view 
of $35\arcsec \times 35\arcsec$ while row (b) shows the same snapshots after 
the subtraction of the average ALMA image. Rows (c), (d), (e), and (f):  same 
as rows (a) and (b) but for the 304 \AA\ data (rows (c), (d)), and the 1600 
\AA\ data (rows (e) and (f)). The white arrows indicate the transient 
brightening in the ALMA and 304 \AA\ data. The bottom row shows time profiles 
of the event emission at 3 mm (black curves) and 304 \AA\ (blue curves), before 
(left panel) and after (right panel) our processing; the processed light curves
show values above the background level. The 304 \AA\ light curves are in
arbitrary units, normalized to fit the vertical extent of the ALMA plots.} 
\end{figure*}

\begin{figure*}[t]
\centering
\includegraphics[width=1.00\textwidth]{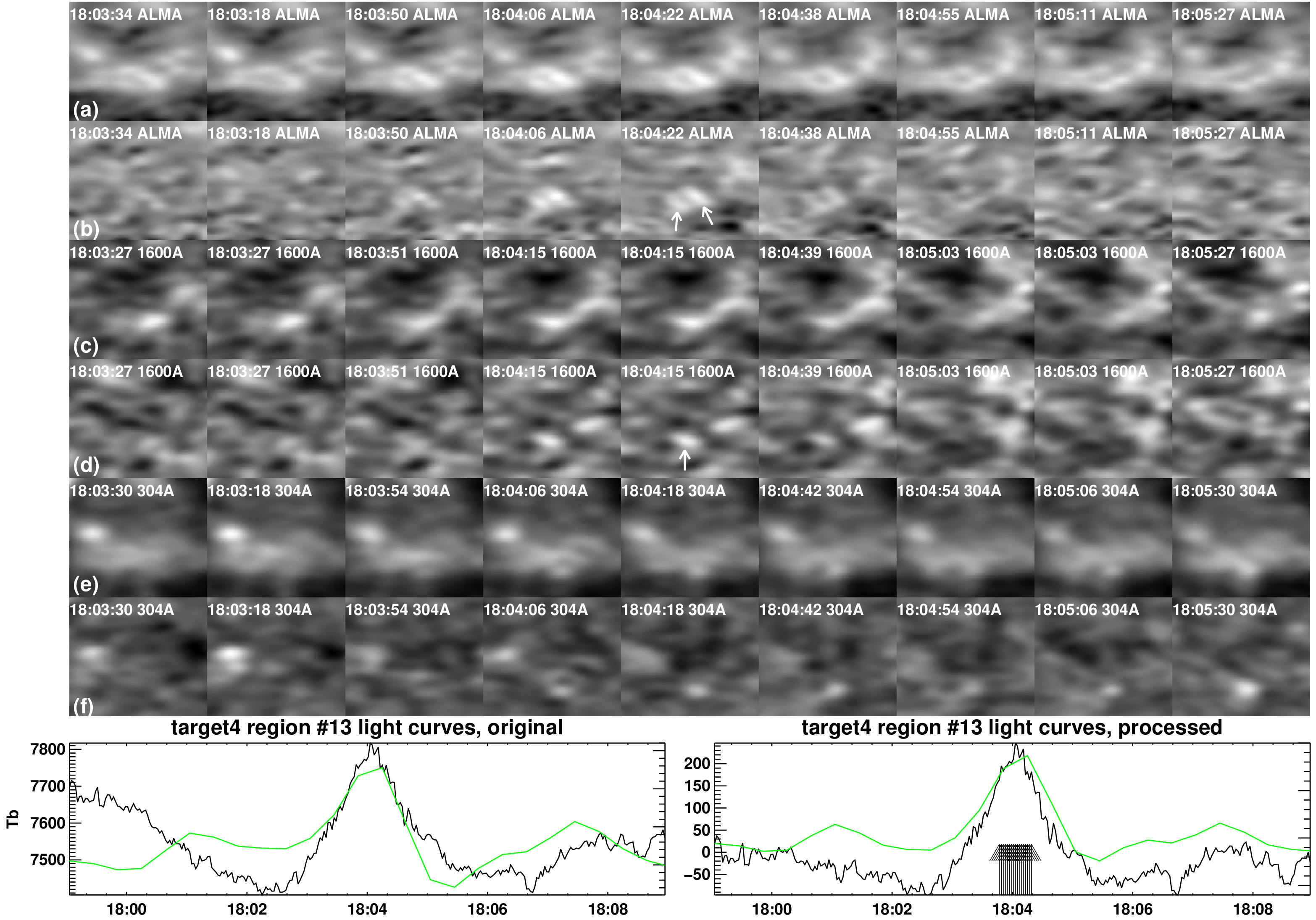}
\caption{Same as Fig. 7 for an event that was detected both in 3 mm and 1600
\AA\ data. The layout of the figure is the same as that of Fig. 7 with the
following exceptions: rows (c) and (d) correspond to 1600 \AA\ data while
rows (e) and (f) correspond to 304 \AA\ data. The white arrows indicate the event
at 3 mm and 1600 \AA. In the bottom row the green curves show light curves from
the 1600 \AA\ data; these light curves are in arbitrary units, normalized to 
fit the vertical extent of the ALMA plots.} 
\end{figure*}

\begin{figure*}[t]
\centering
\includegraphics[width=1.00\textwidth]{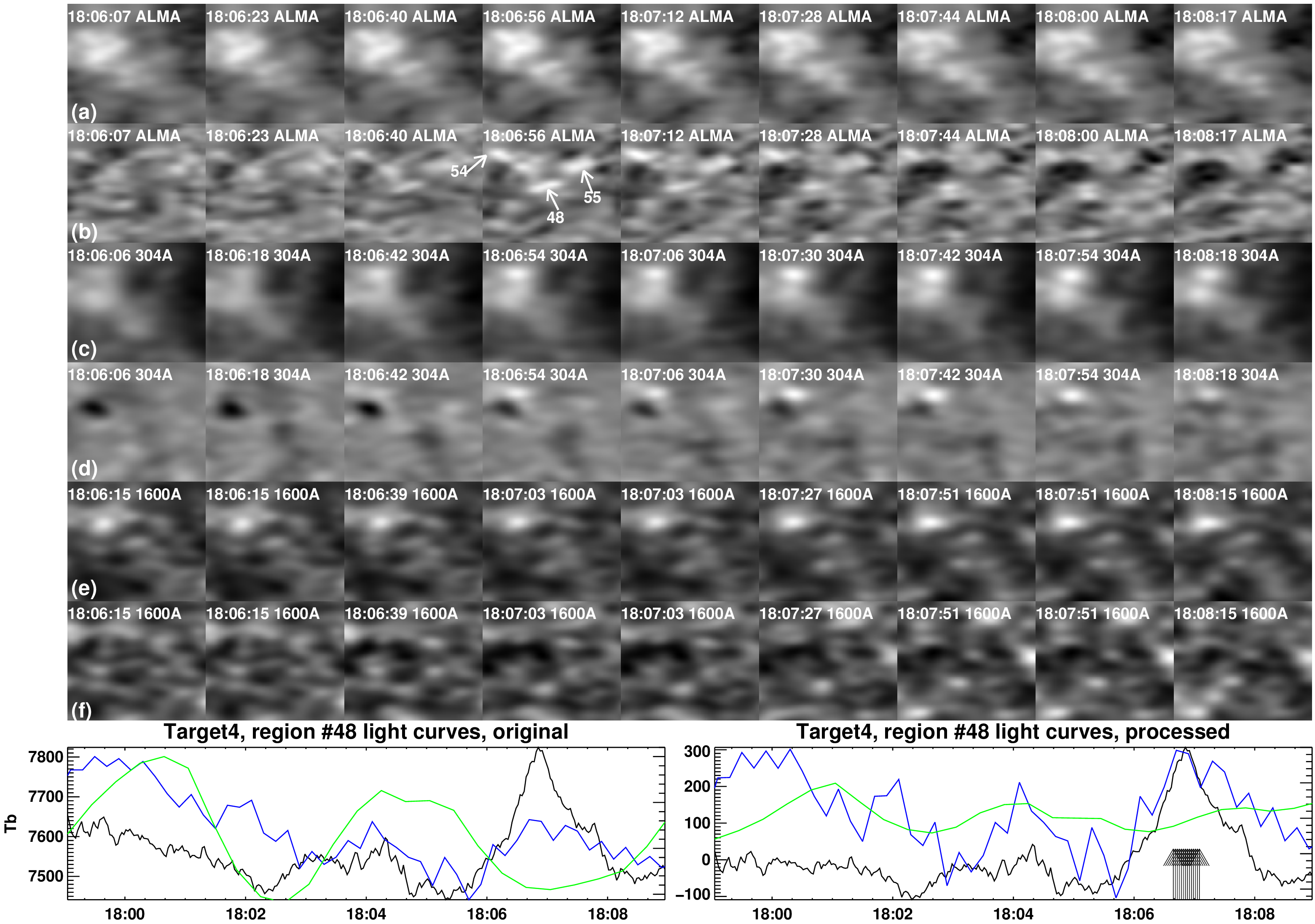}
\caption{Same as Fig. 7 for an event (number 48) that was detected in 3 mm 
data but did not show any conspicuous signature in AIA data. The layout of the 
figure is the same as that of Fig. 7 with the exception that in the bottom 
row the blue and green curves represent light curves from 304 \AA\ and 
1600 \AA\ data, respectively, calculated from the pixels that correspond to 
the ALMA transient brightening.} 
\end{figure*}

In Fig. 5 the pixels that participate in the events detected in target 4 are
indicated with different colors. A similar picture holds for the other targets as well. The top 
row of the figure shows the events detected independently at a given database,
while in the bottom row only the pairs of ALMA-304 \AA\ events and ALMA-1600
\AA\ events are indicated. In all three datasets the detected events show 
remarkable spatial coherence in the sense that they appear as 
irregularly shaped blobs. 

In each snapshot image, we were able to segregate the network from the 
intra-network pixels using the method described by Nindos et al. (2018); that is, 
fit the image intensities with a second-degree polynomial and assume that the 
values above the fit correspond to the network. We found that the ALMA
and 304 \AA\ events show a weak tendency ($\sim 68$-69\%) of appearing at network 
boundaries rather than in cell interiors (see Table 2), which confirms the
visual impression from Fig. 5. The situation is somehow different at 1600 \AA,\ 
where a larger portion of events ($\sim 57\%$) are not associated with network 
boundaries. This may indicate the stronger presence of oscillatory power in 
these images which was not removed adequately. In all three datasets the
percentage of event pixels associated with network boundaries slightly 
increases if we consider only the paired ALMA-AIA events (see Table 2).

The results from our statistical analysis (mean values and frequency
distributions) of the event maximum areas and durations
appear in Table 2 and in Fig. 6. For the ALMA events our calculations
were done twice: first for all events and then only for those that had
AIA  counterparts (either in 304 or 1600 \AA). The properties of the
AIA events  at a given passband that had ALMA counterparts were not
considered separately  because of their small number. The error estimates
associated with the mean values reported in Table 2 were computed
from the standard deviations of the  corresponding distributions. 

The frequency distributions represent number of events per unit of any
given  parameter. We note that all of these distributions can be fitted with
power-law functions above a cutoff. In each case, a linear fit in
logarithmic  axes was performed that gave the power-law function
parameters. The fitting range was determined  automatically as the
widest range of values that yielded mean error bars less than
10\%. Each point within the fitting range was assigned with an
uncertainty computed assuming Poisson statistics of the number of
events in the histogram  bins we used to calculate the frequency
distribution. The propagation of these uncertainties yields
uncertainties on the fitted power-law indices that are reported in
Table 2.  

The mean maximum areas of the events lie between 12.3 Mm$^2$ and 15.5
Mm$^2$. We note that these values were derived after we divided the
apparent areas by $\mu$. Our calculations together with their error
bars indicate that the size of event blobs is similar in all three
wavelengths and this result could be attributed to both the smoothing
of the AIA data  with the ALMA beam and the use of the ALMA-beam size
spatial coherence detection criterion for all three datasets.  The
frequency distributions of event maximum  areas were fitted with
power-law functions extending from  6.5 Mm$^2$ to about 75 Mm$^2$. The
power-law index for the ALMA events is higher than the indices for
both the 304  \AA \ and 1600 \AA\ events, and (given the uncertainties
involved) agrees with that for the ALMA events that had AIA
counterparts.  These values lie between those reported by Joulin et
al. (2016) and Berghmans et al. (1998), who used AIA data in several
passbands and 304 \AA\ EIT data, respectively. There are no events
with areas less than about 6.5 Mm$^2$ as a result of the spatial coherence
detection criterion that we used.

We quantified the average duration of the detected events by
measuring their FWHM in the corresponding light curves (see Table
2). The frequency distributions of event durations are also power laws
extending from 8, 12, and 24 s for the 3 mm, 304, and 1600 \AA\ data,
respectively, to about  100 s.  The different low-end cutoffs reflect
the different cadences and event detection schemes used for the three
datasets (see Section 2). Overall the derived  power-law indices are
within the values reported in the literature (e.g., see Joulin et
al. 2016 and references therein). It is also interesting that, again,
the power-law index of the ALMA events that had AIA counterparts
agrees with that of the general population of ALMA events if
uncertainties  are taken into account. We note, however, that the
power-law fittings cannot account for the high-end values of the ALMA duration
frequency distributions (see Fig. 6(a)). The situation changes if we fit
these distributions
%\LEt{Please specify what "them" refers to.} 
with exponential functions (the relevant best-fit functions were of
the form: frequency distribution $\propto \exp (-a \cdot \mbox{duration})$, 
where $a=0.087$ and 0.082 s$^{-1}$  for all and paired 
events, respectively) at the expense of obtaining relatively poor fits for 
the low-end values of the frequency distributions. Overall, the values of 
$\chi^2$ for the exponential fits are about 10\% smaller than those of the 
power-law fits, which implies that the exponential fits have a slight edge over 
the power-law fits in the modeling of the ALMA duration frequency 
distributions.

The light curves of all events (e.g., see top row of Fig. 1, as well as
those in Figs. 7, 8, and 9) are  gradual. Although the lack of ALMA
data at two frequencies does not allow us to compute the spectral
index of the millimeter-wavelength emission of the events, the  gradual rise
and fall of their light curves strongly suggests a thermal origin via
the free-free mechanism. This behavior is in agreement with the
results  obtained by White et al. (1995) for transient brightenings
detected at 17 GHz, but contradicts other studies of microwave data
that reported the presence of nonthermal populations of electrons
(e.g., Gary et al. 1997; Krucker et al. 1997; Nindos et
al. 1999). However, quantitative calculations verify that only
Mega-electron-volt-energy 
%\LEt{Please spell out MeV and other units used without numbers.}
electrons can produce significant gyrosynchrotron emission
at 3 mm  (e.g., White and Kundu 1992).

In Figs. 7, 8, and 9 we present three characteristic events. The first
event was detected in both ALMA and 304 \AA\ data, the second was detected in
both ALMA and 1600 \AA,\ while the third  appeared only in ALMA data
with no compelling signature in the AIA databases. All events are so
weak that they cannot be readily identified in the plain images, but
their visual identification, as unresolved bright kernels, is possible
after the subtraction of the temporally  averaged image from each
snapshot image. In all these figures the event component dominates the
light curves over the residual oscillatory pattern with the exception
of the 1600 \AA\ and 304 \AA\ light curves of Fig. 9; in the latter
case this was natural because these light curves were calculated from
pixels  that did not belong to any detected AIA event. We note that
although a local peak of  the 304 \AA\ light curve of Fig. 9 occurs close
to the time of the 3 mm peak, it does not qualify as an event because it
did not exceed the 2.5$\sigma$ threshold.

\section{Implications for chromospheric heating}

In this section we present our estimates of the thermal energy
supplied to the chromosphere by the ALMA transient brightenings. By doing 
so we certainly miss other potential carriers
of energy, such as flows
and waves, that might also be associated with such events. We assume that that their emission comes from the thermal free-free 
mechanism (see Section 3). We start from the following well-known expression for 
the thermal energy:

\begin{equation}
E = 3N_e k T_e V_{ap}
,\end{equation}                                          
where $T_e$ is the electron temperature, $k$ is the Boltzmann 
constant, and a filling factor of unity has 
been assumed. Assuming, further, that the electron density, $N_e$, 
and apparent volume, $V_{ap}$, do not vary appreciably during the events, 
the extra energy, $\Delta E_i$, supplied to the chromosphere during the time 
interval between two consecutive images, $i-1$ and $i$, is written as

\begin{equation}
\Delta E_i = 3N_e k\,\Delta T_{e,i}\,V_{ap}
,\end{equation}
where $\Delta T_{e,i}= T_{e,i}-T_{e,i-1}$ is the difference of the 
electron temperature. Under the reasonable assumption that the energy 
release occurs during the rise time of the brightening, the total energy 
provided, $E$, is given as
\begin{equation}
E = 3N_e k V_{ap} \sum_{i=1}^{i=i_{max}} \Delta T_{e,i}
,\end{equation}
where $i_{max}$ is the number of the image at which temperature reaches 
its maximum. The sum in eq. (3) is obviously equal to the peak electron 
temperature, $T_{e,max}$ above background, $T_{e,0}$, that is,  $T_{e,max}-T_{e,0}$, 
so that, finally,
\begin{equation}
E = 3N_e k V_{ap} (T_{e,max}-T_{e,0})
.\end{equation}

From the processed light curves of each event we obtained its maximum
excess brightness  temperature above background $T_{b,max}-T_{b,0}$. 
In order to find the $T_{e,max}-T_{e,0}$  which is 
required by eq. (4) we need to 
know the  optical depth, $\tau$. To this end we used the results by 
Alissandrakis et al.  (2017) who, after they inverted the center-to-limb 
variation curve of full-disk data, found electron temperature  values in the 
range 7250-7950 K, which were only 5\% lower than those provided by the Fontenla
et al. (1993) FAL C model. In that model the above electron temperatures 
correspond to heights from $h_1 \approx 1775$ km to $h_2 \approx 1950$ km. If 
we use that range of electron temperatures and the corresponding FAL C values 
for $N_e$, we find that our events occur in configurations in which the optical 
depth ranges from about 10 to 16, that is, they are optically thick. 
Therefore it is reasonable to use the excess brightness temperature above 
background from our corrected 3 mm light curves for the computation of the 
excess electron temperatures involved in eq. (2).  

For the evaluation of the energy budget we used the same FAL C
electron density values as in the calculation of $\tau$. For each event our
calculations were done twice: first for the values of electron density
corresponding to $h_1$ and then for their values that correspond to
$h_2$. 

Finally, we assume that  the
volume, $V_{ap}$, can be estimated from the area, $A$, of the  event
through the equation $V_{ap} = A^{3/2}$, that is, we assume that the
extent in  height is comparable to the horizontal size.  
Our assumptions for constant electron  density and filling factor
$f=1$ imply that the derived energies represent upper limits and as does the assumption about the vertical scale of the events, which
should be comparable to the vertical extent of the chromosphere, an
extent over which the plasma properties may change dramatically.

Our computations (see Table 3) yielded thermal energies ranging
from $(1.5 \pm 0.1)  \times 10^{24}$ to $(9.9 \pm 2.0) 
\times 10^{25}$ erg.  The uncertainties come  from the range 
of electron temperatures
and densities used in each event  calculations and from the root mean 
square (rms)
%\LEt{Please introduce on first reference.} 
of the
distribution of apparent areas of the  events. The lower-end values of
the derived energies is consistent with the high-end limit of the
nominal nanoflare energy (10$^{24}$ erg). Furthermore, the range of
computed energies falls within the cluster of values that have been
reported in the literature: they are consistent with or smaller than
the  values reported by Krucker \& Benz (1998; $8 \times 10^{24}-1.6
\times 10^{26}$  erg), Berghmans et al. (1998; $5 \times 10^{24}-3
\times 10^{27}$ erg), and  Winebarger et al. (2013; $2.0-6.3 \times
10^{24}$ erg). On the other hand, Aschwanden et al. (2000), Parnell
\& Jupp (2000), and Subramanian et al. (2018) reported energy ranges
($5 \times 10^{23}-5 \times 10^{26}$ erg, $10^{23}-10^{26}$ erg, and $0.3-30.0 
\times 10^{24}$ erg, respectively) whose low-end values are
below the lowest energies we detected. Moreover, studies of active
region weak transient brightenings report larger energies than ours
(e.g., Shimizu 1995; $10^{25}-10^{29}$ erg and Hannah et al. 2008;
$10^{26}-10^{30}$ erg).

We also computed the frequency distribution (i.e., number of
events per unit energy) of all ALMA transient brightenings as a
function of their energy.  The results for all ALMA events is given by
the solid black curve of Fig. 10.  The gray band shows the
uncertainties in the frequency distribution which  incorporate the
error bars associated with both the energy calculations and the
construction of the frequency distribution (see Section 3). For
energies higher than 2.4 $\times 10^{24}$ erg (i.e., if we exclude the
extreme low-end part of  the computed energies) the frequency
distribution of events can be fitted with  a power-law function with
index of $1.67 \pm 0.05$ (see the blue curve in Fig.  
10).  The derived power-law index is consistent with indices derived for RHESSI
microflares (1.7; Hannah  et al. 2008), AIA coronal brightenings
(1.65-1.94; Joulin et al. 2016), and  large flares observed by AIA
(1.66; Aschwanden \& Shimizu 2013).

In Fig. 10 the red curve shows the frequency distribution as a
function of  energy of only those ALMA events that had AIA
counterparts (either 304 or 1600  \AA). The minimum energy of that
population was 2.6 $\times 10^{24}$ erg (i.e.,  the low-end energy
values of the general population of events were missing, but the
minimum energy of the population of paired events was
%\LEt{Please specify what "it" refers to if not "the low-end energy values.} 
close to
the cutoff value used for the fitting of the frequency distribution
of the general population of events)
while its maximum energy was very similar to that of the
general population of events. The frequency distribution of the paired
events was fitted with a power-law function with index $1.65 \pm
0.06$,  consistent with the power-law index derived for the general population 
of events.

\begin{figure}[h]
\centering
\includegraphics[width=0.50\textwidth]{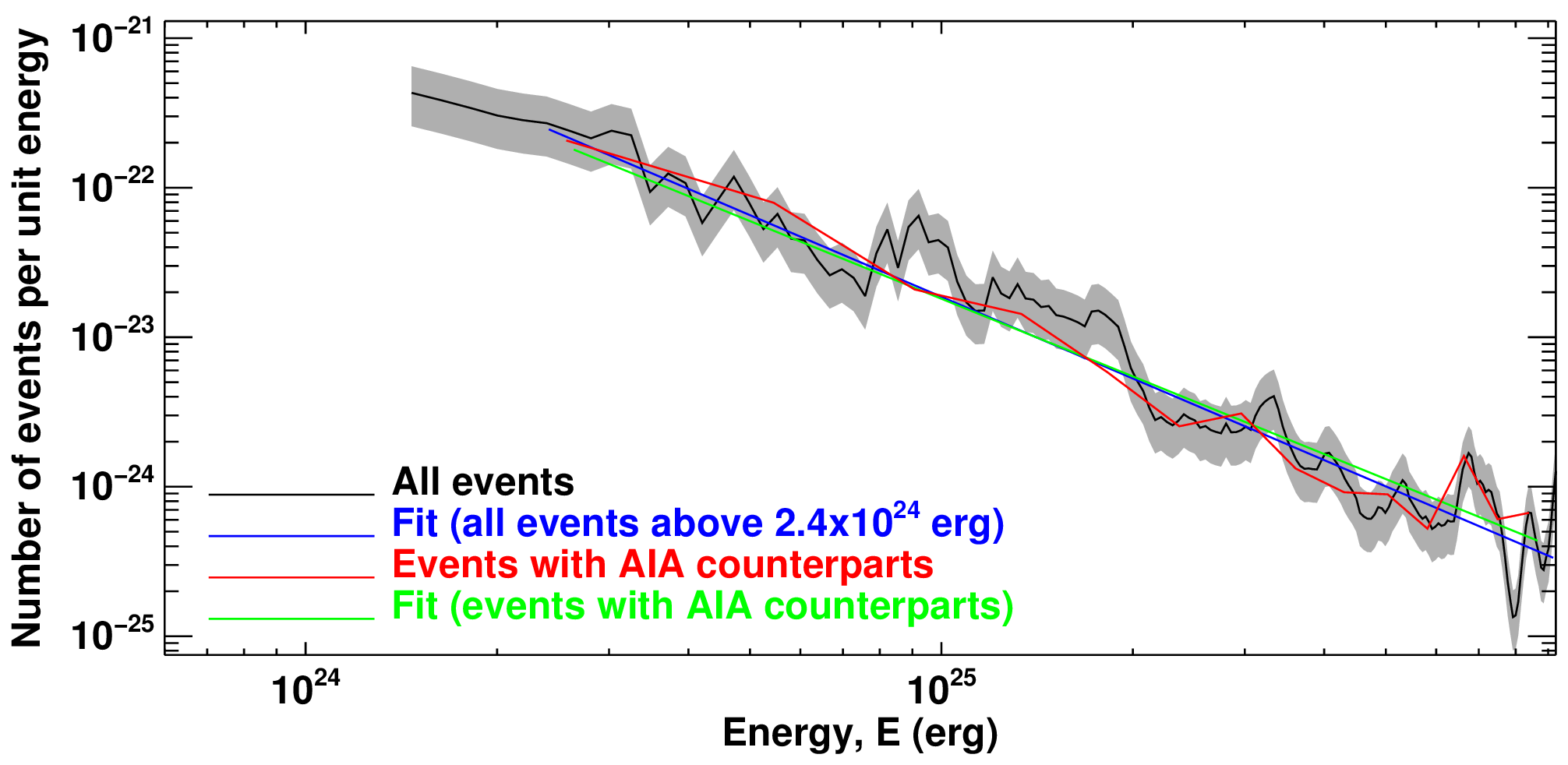}
\caption{Frequency distribution of all ALMA 
transient brightenings as a function of energy shown with black curve. The gray band represents
the error bars discussed in Section 4. For energies $>$ 2.4 
$\times 10^{24}$ erg, the frequency distribution of events has been fitted 
with a power-law function with index of 1.67 (blue curve). The red 
curve corresponds to the frequency distribution vs. energy of only those 
ALMA events that have also been detected in AIA data (either 304 \AA\ or 1600 
\AA). The green curve represents the fitting of that distribution with a 
power-law function with index of 1.65.}  
\end{figure}

\begin{table*}
%{\small
\begin{center}
\caption{Energy budgets of ALMA transient brightenings}
\label{ALMAobs}
\begin{tabular}{lcccc}
\hline 
Population & Minimum energy & Maximum energy & Power-law index & Power per unit area  \\
           & (10$^{23}$ erg) &  (10$^{26}$ erg) & & (10$^4$ erg cm$^{-2}$ s$^{-1}$) \\
\hline

All events (2.5$\sigma$, pixels $\geq $ beam size) & $15.0 \pm 1.0$    & $1.0 \pm 0.2$ & $1.67 \pm 0.05$ & $1.9 \pm 0.2$ \\
Paired events (2.5$\sigma$, pixels $\geq $ beam size) & $26.0 \pm 3.0$ & $0.9 \pm 0.3$ & $1.65 \pm 0.06$ & $1.7 \pm 0.3$ \\
All events (2.3$\sigma$, pixels $\geq $ beam size)  & $7.0 \pm 2.0$  & $1.0 \pm 0.3$ & $1.70 \pm 0.03$  & $2.1 \pm 0.4$ \\
Paired events (2.3$\sigma$, pixels $\geq $ beam size) & $24.0 \pm 3.0$ & $0.9 \pm 0.4$ & $1.67 \pm 0.07$  & $1.9 \pm 0.4$ \\
All events (2.1$\sigma$, $\geq 1$ pixel)  & $4.0 \pm 1.5$            & $1.1 \pm 0.2$ & $1.88 \pm 0.02$  & $3.3 \pm 0.1$ \\
Paired events (2.1$\sigma$, $\geq 1$ pixel) & $23.5 \pm 1.5$          & $0.9 \pm 0.1$ & $1.71 \pm 0.06$   & $2.0 \pm 0.1$ \\
\hline 
\end{tabular}
\tablefoot{In parentheses we give the $\sigma$ multiplication factor and spatial 
coherence criterion used for the detections.}
\end{center}
%}
\end{table*}

From the total amount of energy of the detected events, the duration
of observations in all six targets, and the area of the six fields of
view, we calculated the resulted energy per unit area and time, which
was about 1.9 $\times 10^4$ erg cm$^{-2}$ s$^{-1}$. This
value is factors of 3.8 and 44 smaller than the
power per unit area of the events studied by Krucker \& Benz  (1998)
and Benz \& Krucker (2002), respectively. It is also a factor of
4.6 smaller than the dissipation rate of magnetic energy per
unit area in the  quiet corona computed by Meyer et al. (2013). The
total radiative losses from  the quiet low chromosphere are on the
order of $2 \times 10^6$ erg cm$^{-2}$  s$^{-1}$, which is about one
order of magnitude higher than the relevant quiet  corona losses
(e.g., see Withbroe and Noyes 1977). Therefore the energy supplied by
the weak ALMA transient brightenings can account for only about 1\% 
of the  chromospheric radiative losses and about 10\% of the coronal
radiative losses.

The results of the total thermal energy content of the configurations that
hosted the transient brightenings depend on the event selection
criteria. If less restrictive criteria are adopted, both the  number
and energy content of the selected events increase. For example, we
estimated the energy content of the ALMA events detected by using both
a 2.3$\sigma$ and a 2.1$\sigma$  threshold above the average intensity
for the light curves and by relaxing the spatial coherence criterion
in the latter case. The 2.1$\sigma$ threshold was selected  because
below this threshold, the power-law form of the frequency distribution functions
disappears. The results appear in Table 3. In the least restrictive
case of the 2.1$\sigma$ threshold, the number of detected events
increased by a factor of about 9; the low-end limit of their
energy range dropped  to $(4.0 \pm 1.5) \times  10^{23}$ erg, while 
the power-law index of the frequency distribution of the  events versus
their energy became  $1.88 \pm 0.02$, and their power to unit area
increased to $3.3 \times 10^4$ erg cm$^{-2}$ s$^{-1}$. Interestingly,
the characteristics of the energy distribution of the events with AIA
counterparts did not change very much (see Table 3). 

% 45200 erg s^(-1) cm^(-2) = 45.2 W m^(-2)

\section{Summary and conclusions}

In this article we present the first systematic survey for transient
brightenings in the quiet Sun using ALMA observations at 3 mm. Compared to 
the more usual EUV/soft X-ray (SXR) surveys, the ALMA data have the advantage 
of  the superior cadence and the easier derivation of the physical  properties 
of the detected events. Furthermore, they probe cooler and denser 
chromospheric plasma, which is not accessible with EUV/SXR observations.  

On the other hand, any attempt to search ALMA 3 mm data for transient 
brightenings needs to confront the ubiquitous presence of the p-mode 
oscillations, which could exhibit amplitudes as high as 350 K in individual 
pixels (Patsourakos et al.  2020). To this end, an important component of our
event detection algorithm was the identification and removal of oscillations
from the light curves of individual pixels. There was a slight increase
of residual oscillation power toward the limb, and thus there is no surprise 
that we detected a weak increase of the maximum brightness temperature of the 
event pixels toward the limb as well.

Using our selection criteria (see Section 2) we were able to able to 
detect 184 events in the six $87\arcsec \times 87\arcsec$ targets, each one 
observed for 10 min. All events were of the gradual rise and
fall type, strongly suggesting a thermal origin. The average maximum
brightness temperature of the detected events ranged from
about 70 K to more than 500 K above the average intensity. The mean
values of  their maximum area and duration were 12.3 Mm$^2$
and 51.1 s,  respectively, with a weak preference ($\sim 68\%$) occurring at network  boundaries than in cell interiors. The frequency
distributions of both  parameters followed power-law functions with
indices of 2.73 and 2.35, respectively; we note though that an 
exponential function provided a slightly better fit to the frequency 
distribution of duration. These values are broadly consistent with previous 
reports of these quantities from EUV/SXR observations.  

The detection of ALMA transient brightenings was complemented with the
search for transient brightenings in the corresponding AIA data obtained
at 304 and 1600 \AA. We detected 199 events in 304 \AA\ and 633 in 1600
\AA. As a consquence of the smoothing of the AIA data with the ALMA beam and the
usage of ALMA beam-size patches of adjacent pixels for the spatial
coherence criterion, the size of the events were similar in all
three wavelengths. Furthermore, there was a  weak preference ($\sim 57\%$) for 
the 1600 \AA\ events to occur in cell interiors, which could imply that 
part of the oscillatory strength was not removed. 

Only a small fraction of ALMA events had 304 \AA\ and 1600 \AA\ counterparts
(18 and 14), respectively. The basic properties of the paired ALMA events
were consistent with those of the general population of ALMA events with
the exception that their energy distribution did not reach the
low-end values of the corresponding distribution of the general population.

Regarding the question of why most 3 mm events were not detected at 304 \AA\  or
1600 \AA, and vice versa, we point out that in addition to differences in
sensitivity, which need to be addressed in a future work, the three datasets
probe different ranges of temperatures (or equivalently the corresponding
emissions form at different heights; see Alissandrakis 2019). The 3 mm events 
probe cool chromospheric material whose temperature may not increase 
sufficiently to give rise to 304 \AA\ emission. On the other hand, the 304 \AA\  or   
1600 \AA\ events may not be strong enough to energize the atmospheric
layers that are probed by ALMA. Furthermore, all three wavelengths should 
be optically thick in the context of our study; therefore it is perhaps not
surprising that ALMA transients, seen higher in the chromosphere, do not
correlate well with those that occur lower down because we do not see down
to those heights at 3 mm.

The thermal energies supplied to the chromosphere by the ALMA events
are between $1.5 \times 10^{24}$ and $9.9 \times 10^{25}$ erg  and their 
frequency distribution versus energy follows a
power-law function with index of 1.67. The ALMA events that had AIA
counterparts lack the low-end energy values of the general population
of  events and follow a power-law distribution with index of 1.65. 
The power per unit area supplied by the ALMA events can account for only
1\% of the chromospheric radiative losses or equivalently 10\% of the coronal 
losses.

Of course any calculation of the energy budget of transient brightenings
is sensitive to the detection criteria that have been employed. In our
case using the less restrictive criteria, we derived results that did not change the basic conclusion that the
energy content of the ALMA transient brightenings is not sufficient to
heat the chromosphere. This is the case even though the
number of detected events increased and the calculated energy range extended
down to somehow lower energies. We note, however, that in light of the fact that the 3 
mm emission should be optically thick, we probably detected transient 
brightenings from a thin layer of the chromosphere of only a few hundred kilometer in 
thickness. We speculate that if we were to add up a truly volumetric sample 
of transients occurring at all heights their contribution to heating might 
increase significantly. 

Our observations were carried out with most compact array
configuration available using ALMA at the cost of inferior spatial resolution; we speculate
that the use of higher spatial resolution ALMA observations could
yield the detection of smaller and energetically weaker events that
could lead to steeper power-law functions for the frequency
distribution of events. Observations at ALMA Band 1 at 7.25 mm
(whenever that will become available for solar observing), where the
impact of oscillations is expected to be smaller, could also facilitate
the detection of more events. Another item for future  research are the
physical mechanisms that cause the transient brightenings that we
detected. Previous publications propose that transient chromospheric
temperature increases that may  be associated with the observed
brightenings in  millimeter wavelengths could result from acoustic (Carlsson
\& Stein 1995; Wedemeyer et al.  2004) or magnetoacoustic shocks
(Rouppe van den Voort et al. 2007). These and possibly other
alternatives should be checked against the properties of the transient
brightenings that we detected.

\begin{acknowledgements}
We thank the referee for his/her comments which led to improvement
of the paper. This paper makes use of the following ALMA data:
ADS/JAO.ALMA2016.1.00572.S. ALMA is a partnership of ESO (representing
its member states), NSF (USA) and NINS (Japan), together with NRC
(Canada) and NSC and ASIAA (Taiwan), and KASI (Republic of Korea), in
cooperation with the Republic of Chile. The Joint ALMA Observatory is
operated by ESO, AUI/NRAO and NAOJ.
\end{acknowledgements}

\end{document}